# Structure and solidification of the $(Fe_{0.75}B_{0.15}Si_{0.1})_{100-x}Ta_x$ (x=0-2) melts: experiment and machine learning

I.V. Sterkhova[1,2*], L.V. Kamaeva[1,2], V.I. Lad`yanov[1], N.M. Chtchelkatchev[2]

[1]Udmurt Federal Research Center, Ural Branch, Russian Academy of Sciences, Izhevsk, 426068, Russia

[2]Vereshchagin Institute for High Pressure Physics, Russian Academy of Sciences, 108840, Troitsk, Moscow, Russia

**Abstract**

Fe-B-Si system is a matrix for synthesis of new functional materials with exceptional magnetic and mechanical properties. Progress in this area is associated with the search for optimal doping conditions. This theoretical and experimental study is aimed to address the influence of Ta alloying on the structure of undercooled $(Fe_{0.75}B_{0.15}Si_{0.1})_{100-x}Ta_x$ (x=0-2) melts, their undercoolability and the processes of structure formation during solidification. Small concentration of Ta complicates standard ab initio and machine learning investigations. We developed a technique for fast and stable training of machine learning interatomic potential (MLIP) in this case and uncovered the structure of undercooled $(Fe_{0.75}B_{0.15}Si_{0.1})_{100-x}Ta_x$ (x=0-2) melts. Molecular dynamic simulations with MLIP showed that at Ta concentration of 1 at.% there is a sharp change in the chemical short-range ordering in the melt associated with a change in the interaction of Ta atoms. This effect leads to a restructuring of the cluster formation in the system. At the same time, our experimental investigation shows that melts with a Ta content of 1 at.% have the greatest tendency to undercoolability. Alloying with Ta promotes the formation of primary crystals of $Fe_2B$, and at a concentration of more than 1.5 at.% Ta, also of FeTaB. Herewith, near 1 at.% Ta, the crystallization of the melt proceeds nontrivially: with the formation of two intermediate metastable phases $Fe_3B$ and $Fe_2Ta$ Laves phase. Also, the highest tendency to amorphization under conditions of quick quenching is exhibited by a melt with a Ta concentration of 1 at.%. The results not only provide understanding of optimal alloying of Fe-B-Si materials but also promote a machine learning method for numerical design of metallic alloys with a small dopant concentration.

**Key words**:

Fe-B-Si-Ta intermetallics, crystallization, undercoolability, microstructure, solidification

**Introduction**

---

* Corresponding author: irina.sterkhova@mail.ru

The Fe-B-Si system is the basis of a large number of industrially important amorphous and nanocrystalline materials with high mechanical and magnetic properties [1-4]. Increasing the amorphizing ability of these alloys, i.e. increasing the thickness of the amorphous phase is an urgent task. Alloying of alloys of this system with refractory elements such as Nb, Hf, Zr, Ta increases their glass-forming ability (GFA) and the possibility of obtaining bulk amorphous materials [5-9]. Recently, studies of the influence of such additions on the amorphizing ability and properties of these alloys have been actively conducted [8-11]. For increase of the GFA of alloys, it is important both to choose the composition of the alloy and to determine the optimal conditions for their production from the liquid phase (quenching temperature, thermal treatment of the melt) [12, 13]. To select alloy compositions and optimal quenching conditions that provide the best amorphizing ability, one can use an analysis of the temperature and concentration dependences of melt viscosity, undercoolability and features of structure formation during solidification in a wide range of cooling rates. As we have shown earlier [12, 14], this approach makes it possible to effectively predict the optimal quenching conditions for obtaining alloys with the highest proportion of the amorphous phase.

Previously, in [15], we investigated the effect of Ta on the kinematic viscosity of Fe-B-Si melts. The conducted studies have shown that alloying with Ta does not have a significant effect on the viscosity of Fe-B-Si melts. However, on the concentration dependences of the viscosity of the $(Fe_{0.75}B_{0.15}Si_{0.1})_{100-x}Ta_x$ (x=0-2) melts, a local maximum was found near 1 at.% Ta, which indicates a change in the short-range ordering in the liquid phase. A change in the state of the liquid phase will affect the processes of structure formation during solidification, including amorphization.

In this regard, here we study the effect of Ta alloying on the structure of Fe-B-Si melts and the processes of their solidification in a wide range of cooling rates (1-$10^6$ K/s).

**Simulations**

To analyze the structure of undercooled Fe-B-Si-Ta melts near 1 at.% Ta we used machine learning with deep neural networks and *ab initio* methods.

Ab initio part of our investigations involved Vienna ab initio Simulation Package (VASP) [16, 17] with projector augmented wave potential (PAW-PBE) [18] as well as generalized gradient approximation (GGA) for the exchange and correlation part of the DFT potential. The energy cut off 500 eV was selected with energy convergence at $10^{-6}$ eV. Elementary cells consisted of 512 atoms with periodic boundary conditions. Brillouin zone was sampled at GAMMA point.

The theoretical study of melts involves the use of molecular dynamics. Low concentrations of Ta atoms complicate calculations by the *ab initio* quantum molecular dynamics (QMD), since the cells for modeling are quite small and, in addition, it is necessary to integrate the system dynamics for a rather long time (see Fig. 1). So, to improve sampling of configurational space 10 independent QMD trajectories were developed with different initial random arrangement of atoms. The time step in our QMD was 1fs and system dynamics was sampled typically for 10ps for each independent trajectory.

Machine learning allows to take a step forward. This method allows effectively approximate potential energy manyfold of the system under consideration with *ab initio* accuracy. We built a many-body machine learning interaction potential (MLIP) with the deep neural network core using DEEPMD package [19] and DPGEN concurrent learning platform for the generation of reliable deep learning based potential energy models [20].

Creating MLIP requires "training" and "validation" datasets that contain energies, interatomic forces, and virials for a range of Fe-B-Si-Ta configurations [R.E.Ryltsev and N.M.Chtchelkatchev, J. Mol. Liq. Volume 349, 118181 (2022)]. Training datasets were generated by DPGEN while VASP was used as an *ab initio* engine for calculation of energy, interatomic forces and virials. The validation dataset was prepared on top of QMD simulations.

Training the final MLIP we have used the "hybrid descriptor" consisting of two sub-descriptors [19]. They map atomic configuration (energies, interatomic forces, and virials) to a set of symmetry invariant features. The sub-descriptors were constructed from all the information (both angular and radial) of atomic configurations. However, the first sub-descriptor (se_e2_a) took the distance between atoms as the "highest priority" input while the second one (se_e3) adopted angles between two neighboring atoms, see Ref. [19] and Supplementary Material for details. As the result we got the many body MLIP within the cut-off radius.

MLIP training was a bit tricky because of small concentration of Ta atoms. The neural network contains parameters characterizing the interaction of all atoms, including Ta, with each other within the certain cut-off range (we used 7 Å). At a low concentration of Ta, these atoms relatively rarely approach each other in the melt. But a part of the neural network is reserved for Ta-Ta interaction. So, a training dataset should contain large enough frames with close approaching Ta atoms at a distance less than 7 Å. It is rather problematic to create a sufficient number of such configurations at a low Ta concentration. An insufficient number of such frames leads either to the divergence of the training process of the neural network, or to the effects of "amorphization" of the neural network during its training. We have solved these problems in the general case for any alloy where some of the elements have a low concentration. Quick convergence of the training process was achieved due to the inclusion of the artificial compositions with high concentrations of Ta (up to 15%) in the training datasets. In fact, we included a many fold of other alloys in the dataset, continuously decreasing the concentration of Fe, B and Si and increasing it for Ta.

The final MLIP had the following Root Mean Square Error (RMSE) compared to ab initio: energy RMSE/atom=3.5e-03 eV, force RMSE=2.5e-01 eV/A and virial RMSE/atom=1.2e-02 eV.

Fig.1 shows results of MLIP accuracy test. We took 512 atoms in the cell and compared *ab initio* and classical MD. As follows MLIP has *ab initio* accuracy. Fig.1c illustrates that classical molecular dynamics with neural networks allows investigating effect of Ta on local order of the melt while QMD does not allow that because of natural computational limitations on the quantum molecular dynamic trajectory length.

Most results discussed below for Fe-B-Si-Ta melts were obtained using classical molecular dynamics with MLIP as the interaction potential, taking 5000 atoms in a cell and 100ps long molecular dynamics trajectories. We also always developed (to improve sampling) 10 independent trajectories with different random starting atom configurations for each simulation.

**Experimental methods**

The studied samples with different Ta concentration were obtained by melting the corresponding proportions of $Fe_{75}B_{15}Si_{10}$ and $(Fe_{0.75}B_{0.15}Si_{0.10})_{96}Ta_4$ ligatures in a protective atmosphere of purified helium at temperature of 1650°C for 30 minutes. $Fe_{75}Si_{15}B_{10}$ and $(Fe_{0.75}Si_{0.15}B_{0.1})_{96}Ta_4$ ligatures were synthesized from powders of carbonyl iron (Ultra High Purity 13-2), monocrystalline silicon, amorphous boron and metal Ta in $Al_2O_3$ crucibles in an induction furnace in vacuum at a temperature of 1700°C and a pressure of $5·10^{-3}$ Pa with an isothermal exposure of 30 minutes. To homogenize the chemical composition of the obtained ligatures, additional remelting was carried out in a Tamman furnace in an argon atmosphere at 1700°C for 10 minutes, followed by rapid cooling into a copper mold. The chemical composition of the samples was controlled by atomic emission spectroscopy (ICP) on a Spectroflame spectrometer.

The study of the crystallization processes of $(Fe_{0.75}B_{0.15}Si_{0.1})_{100-x}Ta_x$ (x=0-2) melts was carried out by the method of differential thermal analysis (DTA) on a high-temperature analyzer (BTA 983), the measurement procedure for which is described in detail in [21], X-ray diffraction analysis and metallography. DTA thermograms were obtained in the heating mode at a rate of 20°C/min from 100°C to 1650°C and subsequent cooling at a rate of 100°C/min. From the obtained heating and cooling thermograms, the temperatures of all stages of melting (in the heating mode) and crystallization (in the cooling mode) for each alloy were determined, as well as undercoolability values were calculated as the difference between the liquidus temperature determined from the heating thermogram and the crystallization onset temperature determined from the cooling thermogram. The samples after DTA were subjected to metallographic studies and X-ray diffraction analysis.

Metallographic analysis was carried out using a Neophot 21 optical microscope with digital image processing. The samples were pressed into bakelite and ground on diamond wheels (with particle sizes from 75 μm to 15 μm), polishing was carried out on cloth with an aqueous solution of aluminum oxide (3 and 1 μm). Metallografic sections were etched with a 2% alcohol solution of picric acid.

X-ray diffraction analysis of the obtained alloys was carried out on a diffractometer on a θ-θ diffractometer D8 Advance (Bruker AXS) in a parallel beam geometry with Cu-Kα radiation, a Goebel parabolic mirror (60 mm) on the primary beam, and a Sol-XE semiconductor Si(Li) detector.

**Simulation results**

We studied the structure of undercooled $(Fe_{0.75}B_{0.15}Si_{0.1})$-Ta melts near 1 at.% Ta by machine learning with deep neural networks and *ab initio* methods. (In this range of Ta concentrations, a maximum is observed on the concentration dependences of viscosity.) Total and partial pair radial distribution functions (RDFs) were constructed and the main characteristics of the melt local structure were determined: the distances between the nearest neighbors as well as the coordination numbers, which are given in Table 1. Based on these data, we found the Warren-Cowley parameters, which allow us to analyze the chemical short-range order [22, 23]. The total RDFs of the three investigated alloys are practically the same, so Fig. 2a shows the total RDF only for the undercooled $(Fe_{0.75}B_{0.15}Si_{0.1})_{99.5}Ta_{0.5}$ melt. Despite such a low concentration of Ta, the shape of RDF reflects the main features of the total RDFs for the studied melts. Fig. 2a shows that the first peak has two arms near the radius of the first coordination sphere: more pronounced for r smaller than the radius of the first coordination sphere ($R_1$) and less pronounced for $r>R_1$. The formation of the arm at $r<R_1$ is associated with the position of the first maxima of the partial RDFs: Fe-B, B-B and Si-B (Fig. 2 a, b, c, d) and at $r>R_1$– the interaction with the atoms of the alloying element Ta (Fe-Ta, Si-Ta, Ta-Fe, Ta-Si) (Fig. 2 a, b, d, e). For partial RDFs with B and Si, a splitting of the second maximum is observed (Fig. 2c, d).

This type of RDF indicates the complex nature of the short-range order in the studied melts. The introduction of Ta into the Fe-B-Si melt leads to an increase in the range of interaction radii of the components, and pair interactions of identical radii are formed. An increase in the concentration of Ta in the melt manifests itself in pair partial radial distribution functions with Ta (Fig. 2 f, g), the Ta-Ta interaction changes most significantly.

The distribution histogram the Warren - Cowley parameters for the $(Fe_{0.75}B_{0.15}Si_{0.1})_{99}Ta_1$ melt is shown in Fig. 2 h, i. The Warren-Cowley parameters calculated for the three studied compositions with Ta concentrations of 0.5, 1, and 1.5 at.% and the most pronounced concentration changes in the pair interactions of atoms are shown in Fig. 3. According to the results obtained, a pronounced chemical short-range order is observed in the studied melts. The strongest chemical interactions are observed around B and Si atoms. The effective repulsion of B and Si atoms, both from each other and from their own kind, leads to the fact that Fe and Ta metal atoms predominate in the environment of these atoms. The environment of Fe atoms is determined by the well-known effective attraction of Fe and Si, as well as by the strong effective attraction between iron and Ta atoms. The environment around Ta atoms is determined by the composition of the alloy, the largest differences are associated with the interaction of Ta atoms with each other. It can be seen that with an increase in the Ta concentration to more than 1 at.% in the melt, the number of Ta atoms near the Ta atoms becomes less than the statistical average; effective attraction between Ta atoms transforms into effective repulsion. In addition, from the presented data (Fig. 3) it can be seen that the values of the Warren-Cowley parameter for most interactions in the melt with 0.5 at.% Ta differ significantly compared to two neighboring compositions, and for 1 at.% Ta their maximum or minimum values are observed. The observed feature also manifests itself in other characteristics of the short-range order in the studied melts, for example, in the radius of the first coordination sphere (Table 1).

The method of rotational invariants was used for the geometric analysis of the structure of the studied melts [24-26]. Presented in Fig. 4 distributions were compared with the values of rotational invariants for the main types of close-packed structures, which have the following values: for fcc: $W_6$ = -0.01316, $q_6$ = 0.5745; for hcp: $W_6$ = -0.01244, $q_6$ = 0.4847; for the icosahedral phase (ico): $W_6$ = -0.1697, $q_6$ = 0.6633 [24-26]. The analysis performed showed that the formation of two types of clusters, fcc and ico, is typical for the studied melts in a undercooled state. At the minimum Ta concentrations, all types of atoms participate in the formation of ico clusters. An increase in Ta concentration leads to the fact that ico-clusters cease to form around Ta atoms.

Thus, an analysis of the structure of undercooled $(Fe_{0.75}B_{0.15}Si_{0.1})$-Ta melts showed that at a Ta concentration of 0.5 - 1 at.%, there is a sharp change in the chemical short-range ordering in the melt, which is associated with a change in the interaction of Ta atoms, which leads to change in the nature of cluster formation in the system and should be reflected in the processes of structure formation during the melt solidification.

**Experimental results**

Characteristic cooling thermograms of the studied $(Fe_{0.75}B_{0.15}Si_{0.1})_{100-x}Ta_x$ (x=0-2) alloys are shown in Fig. 5. It can be seen that the crystallization of the studied melts has a multistage character. At the same time, with an increase in the concentration of the alloying element in the alloy, a change in the nature of crystallization is observed, and a more significant change in the type of thermograms is observed for melts with a Ta concentration of more than 1 at.%. In this case, a clearly distinguished first stage of crystallization appears on the cooling thermograms (Fig. 5).

The concentration dependence of the undercoolability value under which the crystallization of the melts began is shown in Fig. 6a and has a nonmonotonic character. When Ta is alloyed up to 2 at.%, with an increase in the concentration of the alloying element, the magnitude of undercoolability decreases slightly. In this case, a local maximum is observed near 1 at.% Ta. It should be noted that the nature of the concentration dependences of the undercoolability value of the studied melts repeats the form of the concentration dependences of the kinematic viscosity of these melts [15]: nonmonotonic changes are observed in the region of 1 at.% Ta (Fig. 6b). At that the liquidus line in the studied concentration range, plotted from heating thermograms, practically does not change with an increase in the concentration of the alloying element (Fig. 6c).

The observed features of cooling thermograms and the concentration dependence of undercoolability indicate a change in the nature of structure formation. In this regard, to study the features of the processes of structure formation, metallographic studies of these alloys obtained by cooling from 1650°C at a rate of 100°C/min were carried out.

The metallographic analysis of the ingots also indicates a complex multiphase crystallization of the studied melts. The microstructure of the base alloy, $Fe_{75}B_{15}Si_{10}$, is quite homogeneous and is represented by a large amount of the structure of the eutectoid decomposition of the $Fe_3B$ boride, which crystallizes first from the melt, and a small fraction of the equilibrium eutectic ($Fe+Fe_2B$) (Fig. 7a). An increase in the Ta concentration promotes the nucleation of $Fe_2B$ crystals. In the structure of the alloy ingot with 1 at.% Ta, large $Fe_2B$ dendrites are clearly visible (Fig. 7b). In this case, along with the formation of the metastable $Fe_3B$ boride, which is characteristic of the Fe-B-Si ternary system in the selected concentration range, crystallization of the $(Fe_{0.75}B_{0.15}Si_{0.1})_{99}Ta_1$ melt proceeds with the formation of another metastable phase, the $Fe_2Ta$ Laves phase. We observe traces of the decomposition of these phases in the microstructure of the ingots (Fig. 7 c, d). A further increase in the Ta concentration is accompanied by the appearance of FeTaB crystals. A small amount of them is present in an ingot with 1.5% tantalum (Fig. 8). In this case, the structure of the two-phase component occupying the space between the $Fe_2B$ and FeTaB crystals changes (Fig. 8). And in an ingot with 2 at.% Ta, the number of FeTaB crystals noticeably increases (Fig. 8). The morphology of these crystals indicates that they nucleate directly from the melt.

The metallographic analysis of ingots of Fe-B-Si alloys alloyed with Ta up to 2 at.% showed that an increase in the concentration of Ta in the alloy is accompanied by a change in the conditions for the nucleation and growth of the $Fe_3B$ boride. In this case, for an alloy with 1 at.% Ta the crystallization ability of the $Fe_3B$ boride decreases, and another metastable $Fe_2Ta$ phase is formed. This conclusion is in good agreement with the results obtained in ab-initio molecular dynamics in the study of rotational invariants.

Thus, the studies performed have shown that during the crystallization of a melt with 1 at.% Ta competition at nucleation of two phases is observed. Such a change in the process of structure formation with a change in the Ta concentration in the alloy, together with the data on viscosity, undercoolability and the results of a structural analysis of undercooled melts, indicates that in the system under study, the alloy near 1 at% Ta will exhibit the best tendency to bulk amorphization.

To study the tendency of these melts to amorphization rapidly quenched ribbons were obtained by the method of melt spinning on a rapidly rotating copper disk from various quenching temperatures. The cooling rate varied from $10^5$ to $10^6$ K/s, while the thickness of the ribbons varied from 40 to 120 μm. An analysis of the rapidly quenched ribbons obtained showed that completely X-ray amorphous ribbons were obtained for the $(Fe_{0.75}B_{0.15}Si_{0.1})_{99}Ta_1$ alloy during its quenching from 1600°C; the thickness of the resulting ribbons was ~40 μm.

**Discussion**

The conducted studies of rapidly quenched rods and ribbons showed that the introduction of Ta increases the tendency of Fe-B-Si alloys to amorphization, but does not lead to their complete bulk amorphization. The $(Fe_{0.75}B_{0.15}Si_{0.1})_{100-x}Ta_x$ (x=0-2) alloys begin to amorphize at cooling rates of $10^4$ K/s. Complete amorphization is observed only for the $(Fe_{0.75}B_{0.15}Si_{0.1})_{99}Ta_1$ alloy upon ultrafast quenching ($10^5$ K/s) from 1600°C. Higher amorphization ability of the alloy alloying with 1 at. % Ta, in comparison with the other investigated alloys (0.5; 0.75; 1.25; 1.5 and 2 at.%) is associated with the features of the structural state of this melt before solidification. Using innovative approaches in modeling the structure of undercooled Fe-B-Si melts alloying with small additions of Ta (up to 2 at.%) by machine learning with deep neural networks and ab initio methods, we were able to show that tantalum does not change the main features of the structure and interatomic interaction in the Fe-B-Si melt, which ensure its high amorphization. These include the small radius of Fe-B interatomic interaction and the effective attraction between Fe and Si atoms. However, the introduction of Ta, the ionic radius of which exceeds the radii of Fe and Si, in small concentrations leads to an increase in the range of interaction radii of the components, which contributes to an increase in the amorphization of these melts compared to Fe-B-Si.

In the system under study, the concentration change of most pair interactions is observed at 1 at.% Ta. So near 1 at.% Ta, there is a significant difference in the Warren-Cowley parameters, compared with alloys alloying 0.5 at.% Ta, for the following interactions Fe-Si, Fe-Ta, B-Ta, Si-Fe, Si-B, Si -Si, Si-Ta, Ta-B, Ta-Si. Moreover, at 1 at.% Ta the maximum effective repulsion between the Ta atoms is observed.

When analyzing the geometric arrangement of atoms using rotational invariants, it was revealed that at 1 at.% Ta, the cluster formation processes change, in which Ta atoms cease to be the centers for the formation of icosahedral clusters, but continue to contribute to the formation of crystal structures. The results obtained showed a good agreement between the change in the tendency to cluster formation in the system under study during modeling and the processes of nonequilibrium crystallization, which is observed when $(Fe_{0.75}B_{0.15}Si_{0.1})_{100-x}Ta_x$ (x=0-2) melts are cooled at low rates. At low Ta concentrations, atoms of all kinds are involved in the processes of cluster formation, which leads to the formation of a solid solution based on Fe and $Fe_3B$, $Fe_2B$ borides. For a melt with 1 at.% Ta, the probability of formation of fcc and ico clusters is the same, which corresponds to the features of structure formation. For the melts with 1 at.%Ta, there is a change in the conditions for the nucleation and growth of the $Fe_3B$ boride and the formation of the metastable $Fe_2Ta$ Laves phase. And in the alloy with 1.5 at.% Ta, in which the formation of fcc clusters around Ta dominates in the undercooled state, a stable (equilibrium) FeTaB boride begins to form during crystallization.

This conclusion is well confirmed by the results of the study of rapidly quenched ribbons of $(Fe_{0.75}B_{0.15}Si_{0.1})_{100-x}Ta_x$ (x=0-2) alloys.

The observed changes in the chemical interaction in the Fe-B-Si system upon alloying with small Ta concentrations, as well as the tendency to cluster formation and structure formation processes are in good agreement with the concentration behavior of the undercoolability value. Despite the fact that the crystallization of melts proceeds in a container according to a heterogeneous nucleation mechanism, we observe a maximum at 1 at.% Ta on the concentration dependence of undercoolability. Thus, the results obtained in this work make it possible to use an approach based on the study of the concentration dependences of undercoolability and structure formation processes to select the composition of alloys with the highest tendency to amorphization.

**Conclusions**

We were looking experimentally and theoretically for the optimal Ta-doping of intermetallic compounds based on Fe-B-Si matrix. So, we studied the structure of undercooled $(Fe_{0.75}B_{0.15}Si_{0.1})_{100-x}Ta_x$ (x=0-2) melts, undercoolability and solidification processes in a wide range of cooling rates. For $(Fe_{0.75}B_{0.15}Si_{0.1})_{100-x}Ta_x$ (x=0-2) it was shown that at a Ta concentration of 1 at.%, a sharp change in the chemical short-range order occurs in the melt associated with a modification in the interaction of Ta atoms which leads to a change in the nature of cluster formation in the system. The melts with a Ta concentration of 1 at.% show the greatest tendency to undercoolability. Alloying with Ta promotes the formation of primary crystals of $Fe_2B$, and at a concentration of more than 1.5 at.% Ta, also of FeTaB. In this case, near 1 at.% Ta, the crystallization of the melt proceeds with the formation of two intermediate metastable intermetallic phases $Fe_3B$ and $Fe_2Ta$ Laves phase. It has been established that the melt with a Ta concentration of 1 at.% exhibits the greatest tendency to amorphization under conditions of rapid quenching.

To analyze theoretically the structure of undercooled Fe-B-Si-Ta melts at very small concentrations of Ta we developed a machine learning interaction potential (MLIP) on top of a database prepared with the help of *ab initio* calculations. MLIP training with conventional methods tends to diverge due to the low concentration of Ta and natural lack of training configurations with short range interacting Ta atoms. We have developed a general way to ensure the convergence of the MLIP neural network training process for alloys with a low concentration of one or more components. For Fe-B-Si-Ta system in hand it was achieved by the inclusion of the high entropy compositions with large concentrations of Ta in the training datasets.

This work has been carried out within the framework of the state assignment of the Ministry of Science and Education of Russia. The analysis of atomic short range ordering in melts was carried out with the support of the Russian Science Foundation (grant RNF 18-12-00438).The experimental investigations were performed using equipment of the Shared Use Centre "Centre of Physical and Physicochemical Methods of Analysis and Study of the Properties and Surface Characteristics of Nanostructures, Materials, and Products" UdmFRC UB RAS" supported by Russian Ministry of Science and Education (Grant #RFMEFI62119X0035). Numerical calculations were performed using computing resources of the federal collective usage center Complex for Simulation and Data Processing for Mega-science Facilities at NRC "Kurchatov Institute" (http://ckp.nrcki.ru/) and supercomputers at Joint Supercomputer Center of RAS (JSCC RAS). Part of calculations was carried out on the "Govorun" supercomputer of the Multifunctional Information and Computing Complex, LIT JINR (Dubna).

**References**

1. 1. G. Zhang, H. Ni, Y. Li, T. Liu, A. Wang, H. Zhang, 2022. Fe-based amorphous alloys with superior soft-magnetic properties prepared via smelting reduction of high-phosphorus oolitic iron ore. Intermetallics. 141, 107441. https://doi.org/10.1016/j.intermet.2021.107441.
2. L.L. Pang, A. Inoue, E.N. Zanaeva, F. Wang, A.I. Bazlov, Y. Han, F.L. Kong, S.L. Zhu, R.D. Shull, Nanocrystallization, good soft magnetic properties and ultrahigh mechanical strength for

Fe82-85B13-16Si1Cu1 amorphous alloys, J. Alloy. Comp., 785 (2019) 25-37. https://doi.org/10.1016/j.jallcom.2019.01.150

3. Y. Wu, K. Peng, L. Tang, W. Zhang, Crystallization mechanism of $Fe_{78}Si_{13}B_9$ amorphous alloy induced by ion bombardment, Intermetallics. 91 (2017) 65-69. https://doi.org/10.1016/j.intermet.2017.08.007.

4. C. Dong, A. Inoue, X.H. Wang, F.L. Kong, E.N. Zanaeva, F. Wang, A.I. Bazlov, S.L. Zhu, Q. Li, Soft magnetic properties of $Fe_{82-83}B_{14-15}Si_2C_{0.5-1}$ amorphous alloys with high saturation magnetization above 1.7 T, J. Non-Cryst. Solids. 500 (2018) 173-180. https://doi.org/10.1016/j.jnoncrysol.2018.07.072.

5. Y.X. Geng, Y.M. Wang, J.B. Qiang, G.F. Zhang, C. Dong, O. Tegus, J.Z. Sun, Fe–B–Si–Zr soft magnetic bulk glassy alloys, Intermetallics. 67 (2015) 138-144. https://doi.org/10.1016/j.intermet.2015.08.006.

6. B. Shen, C. Chang, A. Inoue, Formation, ductile deformation behavior and soft-magnetic properties of (Fe,Co,Ni)–B–Si–Nb bulk glassy alloys, Intermetallics. 15 (2007) 9-16. https://doi.org/10.1016/j.intermet.2005.11.037.

7. Y. Geng, Y. Wang, Z. Wang, J. Qiang, H. Wang, C. Dong, O. Tegus, Formation and structure-property correlation of new bulk Fe–B–Si–Hf metallic glasses, Materials & Design, 106, (2016) 69-73. https://doi.org/10.1016/j.matdes.2016.05.102

8. H. Zhang, Z.C. Yan, Q. Chen, Y. Feng, Z.G. Qi, H.Z. Liu, X.Y. Li, W.M. Wang, 2021. Hardness, magnetism and passivation of Fe-Si-B-Nb glasses. J. of Non-Cryst. Solids. 564, 120830. https://doi.org/10.1016/j.jnoncrysol.2021.120830.

9. R. J. Oh, H. Choi-Yim, K. H. Kang, Thermal and Magnetic Properties of the Co-Fe-B-Si-Ta Alloy System for several Fe/Co, Journal of the Korean Physical Society. 69 (12) (2016) 1813-1816. doi: 10.3938/jkps.69.1813

10. W. Li, Y.Z.Yang, C.L. Yao, Z.W. Xie, C.X. Xie, Glass forming ability and magnetic properties of Fe–B–Si–EM (EM=Ti, Zr, Mo, and Hf) alloys with high iron content, J Mater Sci: Mater Electron. 28 (2017) 10218–10223. DOI 10.1007/s10854-017-6788-7

11. J. Torrens-Serra, P. Bruna, M. Stoica, J. Eckert, Glass-forming ability and microstructural evolution of [(Fe0.6Co0.4)0.75Si0.05B0.20]96-xNb4Mx metallic glasses studied by Mossbauer spectroscopy, J. Alloy. Comp., 704 (2017) 748-759. https://doi.org/10.1016/j.jallcom.2017.02.098

12. I.V. Sterkhova, V.I. Lad`yanov, L.V. Kamaeva, N.V. Umnova, P.P. Umnov On the tendency of the Co-, Ni-, and Fe-based melts to the bulk amorphization, Metallurgical and materials transaction A. 47 (2016) 5487-5495. DOI: 10.1007/s11661-016-3693-2

13. R.F. Tournier, M.I. Ojovan, 2022. Multiple Melting Temperatures in Glass-Forming Melts, Sustainability. 14, 2351. https://doi.org/ 10.3390/su14042351

14. V.I. Lad'yanov, I.V. Sterkhova, L.V. Kamaeva, T.R. Chueva, V.V. Molokanov On the solidification of the $Fe_{50}Cr_{15}Mo_{14}C_{15}B_6$ bulk-amorphized alloy, J. Non-Cryst. Solids. 356 (2010) 65-71. https://doi.org/10.1016/j.jnoncrysol.2009.10.011

15. I.V. Sterkhova, L.V. Kamaeva, V.I. Lad'yanov, N.M. Chtchelkatchev, 2021. Role of Ta and Nb alloying elements on the viscosity of Fe-B-Si melts, Journal of Molecular Liquids. 323, 114636. https://doi.org/10.1016/j.molliq.2020.114636

16. G. Kresse, J. Hafner, Ab initio molecular dynamics for liquid metals, Phys. Rev. B. 47 (1993) 558-561, https://doi.org/10.1103/PhysRevB.47.558

17. G. Kresse, J. Hafner, Ab initio molecular-dynamics simulation of the liquid-metal–amorphous-semiconductor transition in germanium, Phys. Rev. B. 49 (1994) 14251-14269. https://doi.org/10.1103/PhysRevB.49.14251

18. G. Kresse, Joubert, From ultrasoft pseudopotentials to the projector augmented-wave method, Phys. Rev. B. 59 (1999) 1758-1775. https://doi.org/10.1103/PhysRevB.59.1758

19. H. Wang, L. Zhang, J. Han, E. Weinan, DeePMD-kit: A deep learning package for many-body potential energy representation and molecular dynamics, Computer Physics Communications. 228 (2018) 178-184. https://doi.org/10.1016/j.cpc.2018.03.016.

20. Y. Zhang, H. Wang, W. Chen, J. Zeng, L. Zhang, H. Wang, E. Weinan, 2020. DP-GEN: A concurrent learning platform for the generation of reliable deep learning based potential energy models, Computer Physics Communications. 253, 107206. https://doi.org/10.1016/j.cpc.2020.107206.

21. I.V. Sterkhova, L.V. Kamaeva, The influence of Si concentration on undercooling of liquid Fe, J. Non-Cryst. Solids. 401 (2014) 250–253. http://dx.doi.org/10.1016/j.jnoncrysol.2014.01.027

22. B. E. Warren, B. L. Averbach, B. W. Roberts, Atomic size effect in the X-ray scattering by alloys, J. Appl. Phys. 22 (1951) 1493–1496. https://doi.org/10.1063/1.1699898

23. J. Wang, X. Li, S. Pan, J. Qin, Mg fragments and Al bonded networks in liquid MgAl alloys, Comput. Mater. Sci., 129 (2017) 115–122. https://doi.org/10.1016/j.commatsci.2016.12.006

24. P. Steinhardt, D. Nelson, M. Ronchetti, Bond-orientational order in liquids and glasses, Phys. Rev. B. 28 (1983) 784-805. https://doi.org/10.1103/PhysRevB.28.784

25. P. R. ten Wolde, M. J. Ruiz-Montero, D. Frenkel, Numerical calculation of the rate of crystal nucleation in a Lennard-Jones system at moderate undercooling, J. Chem. Phys. 104 (1996) 9932-9947. https://doi.org/10.1063/1.471721

26. W. Mickel, S. C. Kapfer, G. E. Schroeder-Turkand, K. Mecke, 2013. Shortcomings of the bond orientational order parameters for the analysis of disordered particulate matter. J. Chem. Phys. 138, 044501. https://doi.org/10.1063/1.4774084

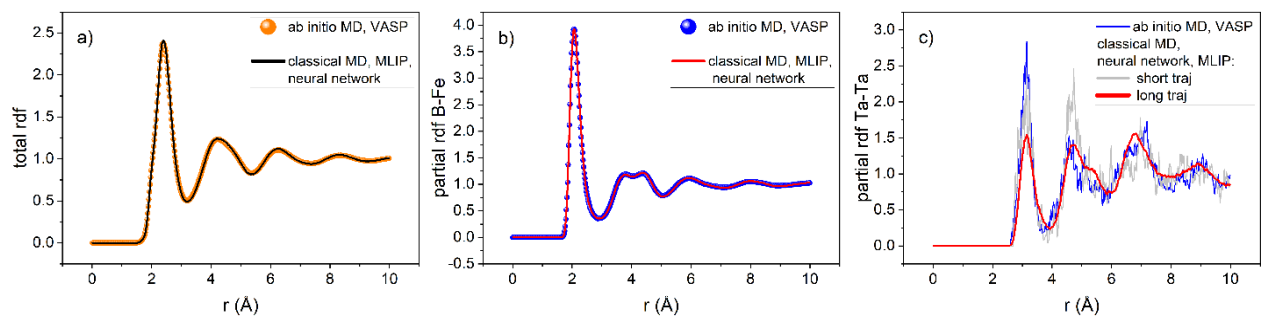

Figure 1. MLIP accuracy test. Radial distribution functions (rdf) of $(Fe_{0.75}B_{0.15}Si_{0.10})_{98.5}Ta_{1.5}$ melt at temperature 1400K and pressure 1 Bar obtained using ab initio quantum molecular dynamics (QMD) and classical molecular dynamics with many body machine learning potential (MLIP) based on deep neural networks. Panel (a) shows total rdf, (b) – partial rdf describing Fe-B, i.e., distribution of Fe around B and (c) partial Ta-Ta rdf where "short traj" means 10ps long QMD trajectories and "long traj" corresponds to 100 ps with MLIP.

| i-j | (Fe$_{0.75}$B$_{0.15}$Si$_{0.1}$)$_{99.5}$Ta$_{0.5}$ | | (Fe$_{0.75}$B$_{0.15}$Si$_{0.1}$)$_{99}$Ta$_{1}$ | | (Fe$_{0.75}$B$_{0.15}$Si$_{0.1}$)$_{98.5}$Ta$_{1.5}$ | |
|---|---|---|---|---|---|---|
| | r$_{ij}$, Å | Z$_{ij}$ | r$_{ij}$, Å | Z$_{ij}$ | r$_{ij}$, Å | Z$_{ij}$ |
| tot-tot | 2.41 | 12.98 | 2.41 | 12.87 | 2.41 | 12.98 |
| Fe-tot | 2.43 | 13.56 | 2.43 | 13.57 | 2.43 | 13.54 |
| Fe-Fe | 2.45 | 10.11 | 2.45 | 10.19 | 2.45 | 10.11 |
| Fe-B | 2.07 | 1.65 | 2.07 | 1.67 | 2.07 | 1.66 |
| Fe-Si | 2.37 | 1.45 | 2.37 | 1.51 | 2.37 | 1.46 |
| Fe-Ta | 2.67 | 0.28 | 2.67 | 0.17 | 2.67 | 0.28 |
| B-tot | 2.07 | 9.10 | 2.07 | 9.01 | 2.07 | 9.08 |
| B-Fe | 2.07 | 8.22 | 2.07 | 8.35 | 2.07 | 8.26 |
| B-B | 1.81 | 0.38 | 1.81 | 0.34 | 1.79 | 0.36 |
| B-Si | 2.25 | 0.29 | 2.25 | 0.27 | 2.23 | 0.27 |
| B-Ta | 2.45 | 0.22 | 2.45 | 0.17 | 2.45 | 0.22 |
| Si-tot | 2.37 | 12.40 | 2.37 | 12.29 | 2.37 | 12.42 |
| Si-Fe | 2.37 | 10.99 | 2.37 | 11.22 | 2.37 | 11.03 |
| Si-B | 2.25 | 0.44 | 2.25 | 0.41 | 2.23 | 0.41 |
| Si-Si | 2.71 | 0.64 | 2.71 | 0.58 | 2.71 | 0.62 |
| Si-Ta | 2.77 | 0.24 | 2.77 | 0.15 | 2.77 | 0.24 |
| Ta-tot | 2.65 | 17.13 | 2.67 | 17.21 | 2.67 | 17.20 |
| Ta-Fe | 2.67 | 13.31 | 2.67 | 13.27 | 2.67 | 13.37 |
| Ta-B | 2.45 | 2.07 | 2.45 | 2.39 | 2.45 | 2.06 |
| Ta-Si | 2.77 | 1.52 | 2.77 | 1.50 | 2.77 | 1.52 |
| Ta-Ta | 3.13 | 0.13 | 3.07 | 0.04 | 3.15 | 0.16 |

Table 1. Distances between nearest neighbors (r$_{i-j}$) and coordination numbers (Z$_{i-j}$) in (Fe$_{0.75}$B$_{0.15}$Si$_{0.10}$)-Ta melts at 1400 K.

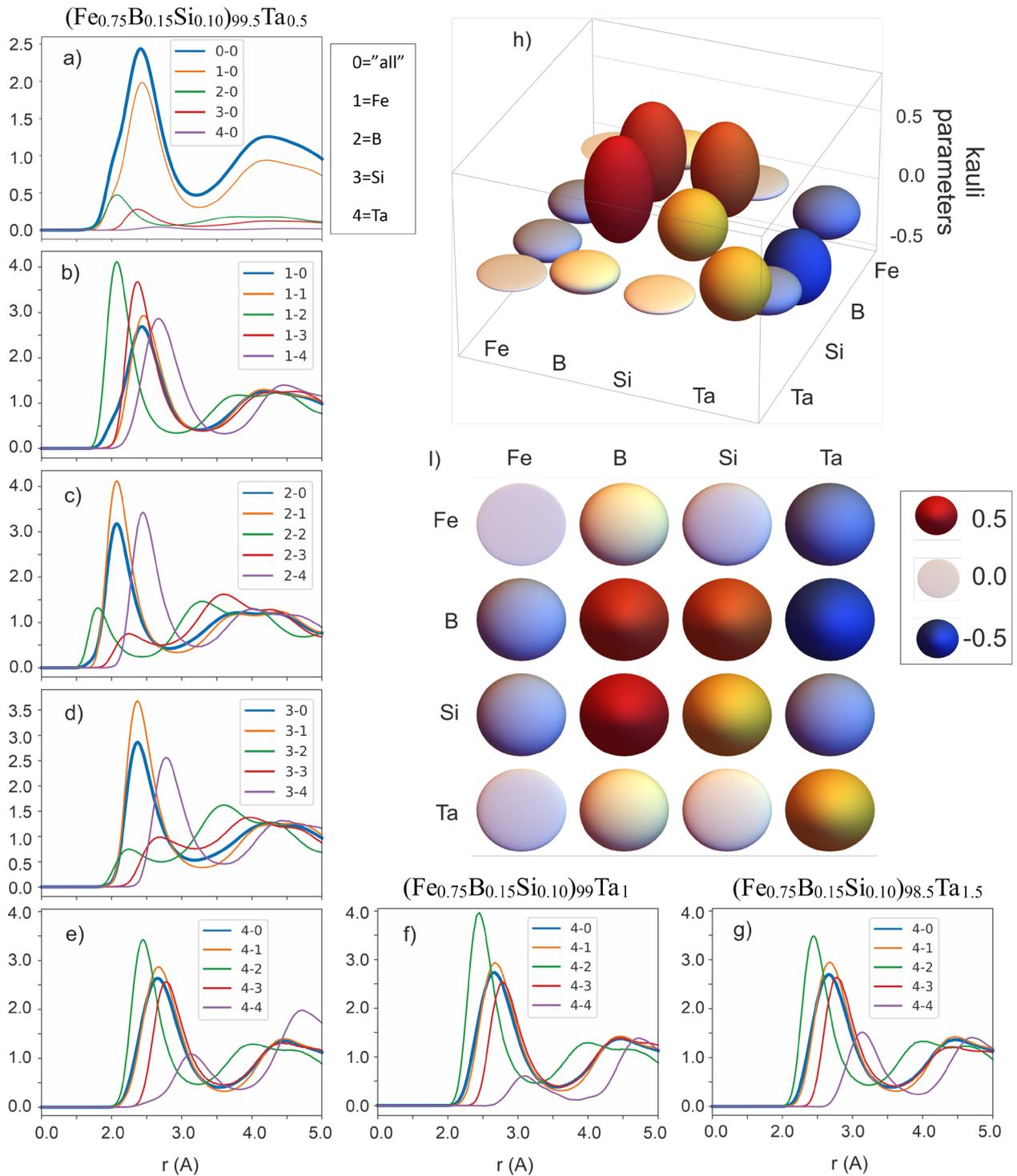

Figure 2. Total and partial RDFs extracted from MLIP simulations for $(Fe_{0.75}B_{0.15}Si_{0.10})_{99.5}Ta_{0.5}$ (a - e), $(Fe_{0.75}B_{0.15}Si_{0.10})_{99}Ta_1$ (f) and $(Fe_{0.75}B_{0.15}Si_{0.10})_{98.5}Ta_{1.5}$ (g) melts at 1400 K and Warren-Cowley SRO parameters for $(Fe_{0.75}B_{0.15}Si_{0.10})_{99}Ta_1$ (h, i).

| at.% Ta | Fe-Fe | Fe-B | Fe-Si | Fe-Ta |
|---|---|---|---|---|
| 0.5 | -0.01 | 0.18 | -0.08 | -3.16 |
| 1 | -0.01 | 0.17 | -0.12 | -0.29 |
| 1.5 | -0.01 | 0.17 | -0.09 | -0.39 |
| | B-Fe | B-B | B-Si | B-Ta |
| 0.5 | -0.21 | 0.72 | 0.68 | -3.78 |
| 1 | -0.25 | 0.74 | 0.69 | -0.74 |
| 1.5 | -0.23 | 0.73 | 0.70 | -0.59 |
| | Si-Fe | Si-B | Si-Si | Si-Ta |
| 0.5 | -0.19 | 0.76 | 0.48 | -2.91 |
| 1 | -0.23 | 0.78 | 0.52 | -0.20 |
| 1.5 | -0.20 | 0.78 | 0.49 | -0.30 |
| | Ta-Fe | Ta-B | Ta-Si | Ta-Ta |
| 0.5 | -0.04 | 0.19 | **0.11** | **-0.49** |
| 1 | -0.04 | 0.07 | **0.12** | **0.79** |
| 1.5 | -0.05 | 0.19 | **0.10** | **0.37** |

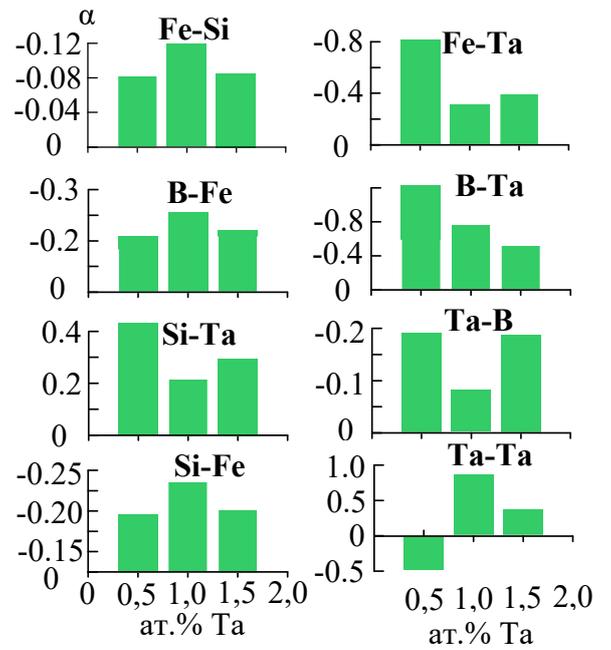

Figure 3. The Warren-Cowley SRO parameters in $(Fe_{0.75}B_{0.15}Si_{0.10})$-Ta melts.

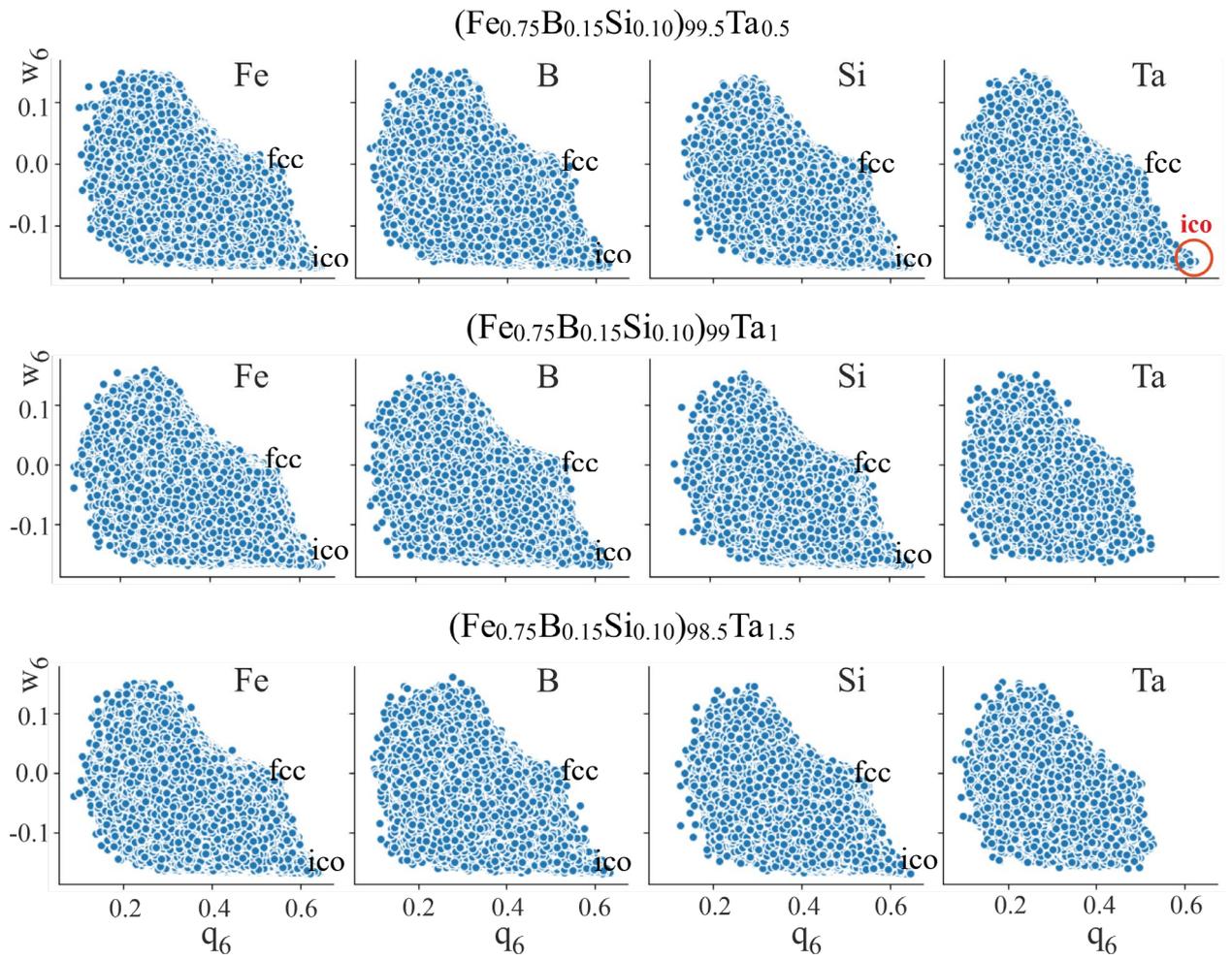

Figure 4. Correlators (joint probability density function) of the bond-orientational order parameters BOOP, $W_6$ and $q_6$, for each atom type in the $(Fe_{0.75}B_{0.15}Si_{0.10})$-Ta melts.

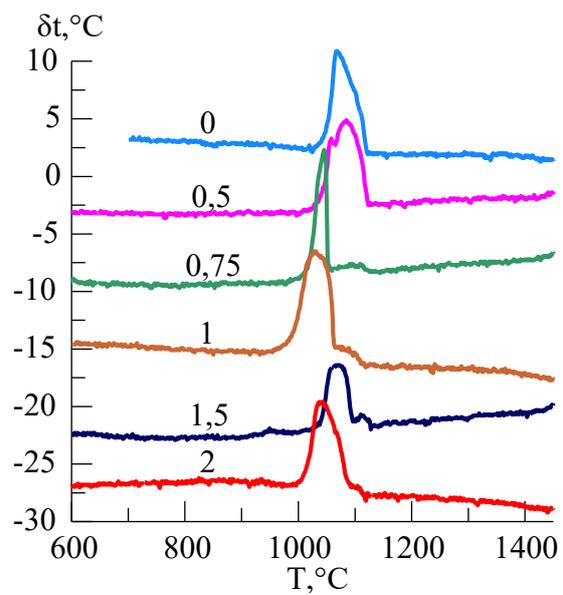

Figure 5. Cooling thermograms of the Fe$_{0.75}$B$_{0.15}$Si$_{0.1}$)$_{100-x}$Ta$_x$ (x=0-2) melts.

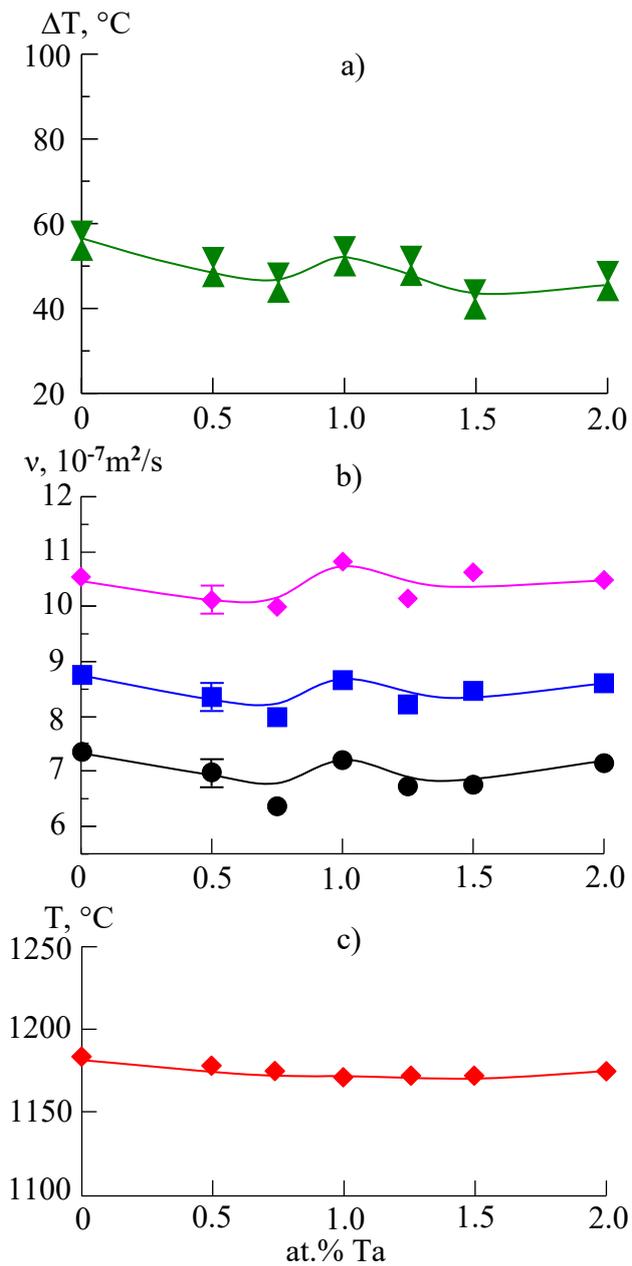

Figure 6. The concentration dependences of undercoolability under cooling from 1650°C at 100°C/min (a), kinematic viscosity (b) at different temperatures: ♦ - 1400°C, ■ - 1500°C, ● - 1600°C [15] and melting points (liquidus temperatures) determined by heating DTA plot at 20 °C/min (c) for $(Fe_{0.75}B_{0.15}Si_{0.10})_{100-x}Ta_x$ (x = 0-2) melts.

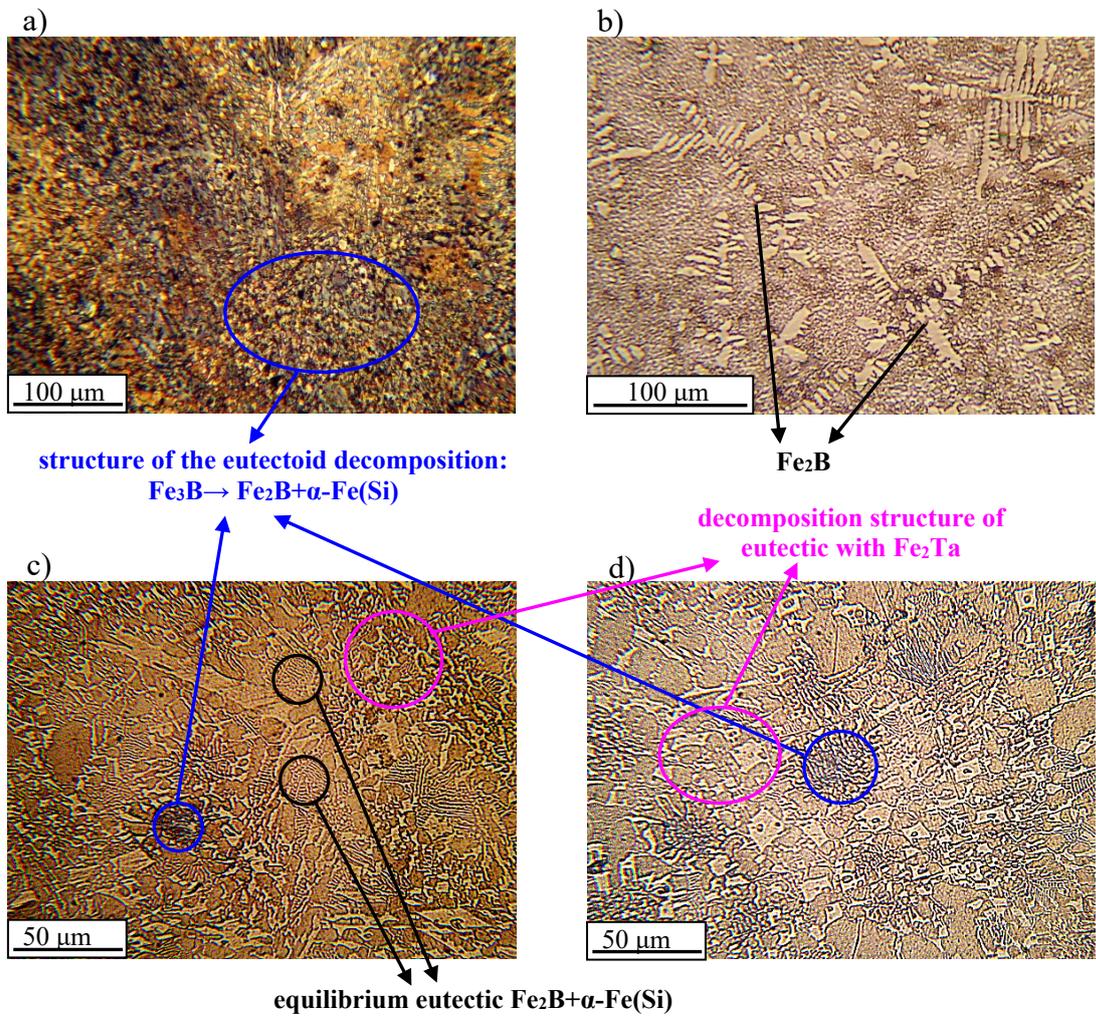

Figure 7. Microstructure of ingots of Fe$_{75}$B$_{15}$Si$_{10}$ (a) and (Fe$_{0.75}$B$_{0.15}$Si$_{0.10}$)$_{99}$Ta$_1$ (b-d) obtained by cooling from 1650°C at 100°C/min.

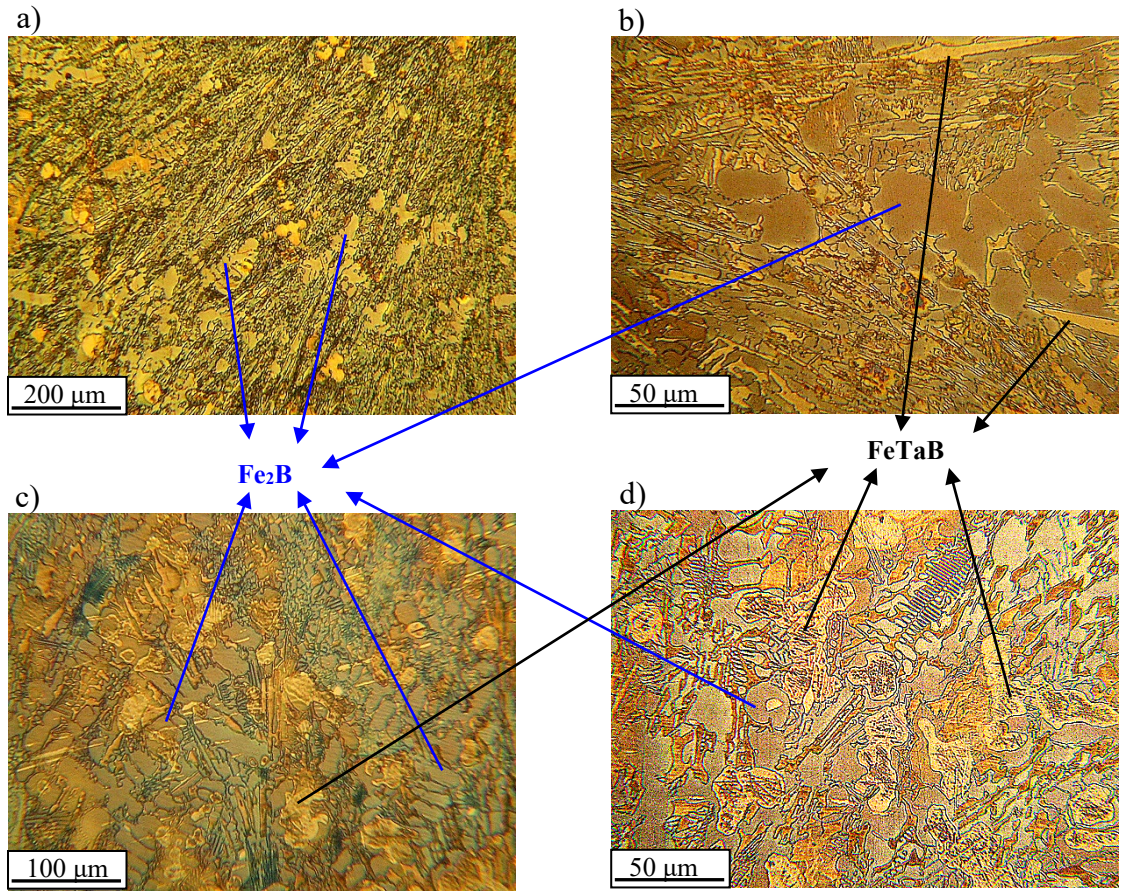

Figure 8. Microstructure of ingots of $(Fe_{0.75}B_{0.15}Si_{0.10})_{98.5}Ta_{1.5}$ (a, b) and $(Fe_{0.75}B_{0.15}Si_{0.10})_{98}Ta_2$ (c, d) obtained by cooling from 1650°C at 100°C/min.

# SUPPLEMENTARY MATERIAL
# Machine learning of neural network interaction potentials
# for doped alloys

When studying alloyed alloys, where not all, but only small concentrations of alloying components are important, the traditional MLIP machine learning procedure usually does not converge (or requires the creation of an inadequately overloaded training database). The fact is that deep learning neural networks contain parameters that characterize the interaction of all types of atoms, including alloying ones. These atoms at low concentrations rarely approach each other in an alloy, but part of the neural network is reserved for describing their interaction and must be trained.

There is a general approach to solving this problem. It is necessary in a certain way to add to the training database a small number of "high-entropy" configurations with a high concentration of alloying elements, where these atoms approach each other with a fairly high probability. This procedure must be implemented carefully, modifying the database "uniformly" along a "continuous path" in the concentration space, starting in the doping region and ending in the high-entropy region, see Fig. 1s.

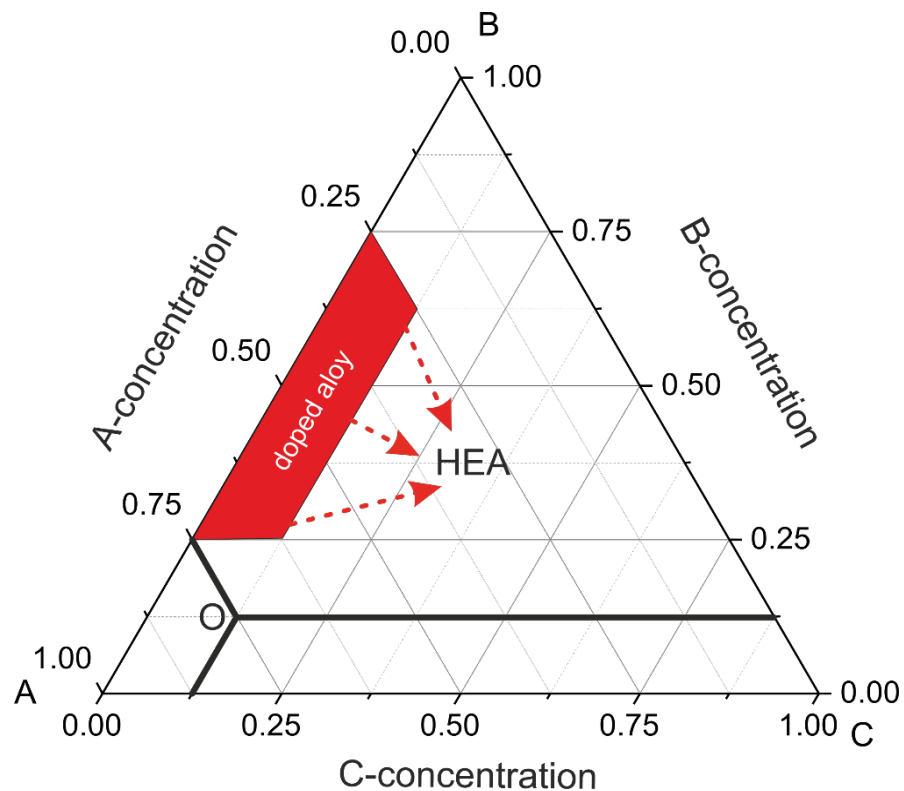

Figure 1s. Trinity chart. To determine the position of the alloy for a given composition, one need to find the content point of the component on the side and draw a line parallel to the other side, opposite the vertex of the triangle for this component. Point O is given as an example. The area of alloyed alloy A-B-C is schematically highlighted in red, where C is the doping component. The arrows towards the high-entropy alloys (HEA) area schematically show the directions in which the database needs to be expanded for MLIP training.

Training the MLIP for FeBSiTa we have used the "hybrid descriptor" consisting of two descriptors: se_e2_a and se_e3_e [1]. They map atomic configuration to a set of symmetry invariant features. The descriptors were constructed from all information (both angular and radial) of atomic configurations. The first descriptor took the distance between atoms as input while the second one adopted angles between two neighboring atoms as input. For se_e2_a descriptor we took rcut_smth=2.0 (A) and rcut=7.0 (A), while for se_e3_e we chose rcut_smth=1.5 (A) and rcut=4.0 (A). The upper cut-off for "angular" descriptor was taken relatively small because interaction between atoms becomes isotropic over long distances. The deep neural network for se_e2_a descriptor consisted of hidden layers with [25,50,100] neurons and with 16 angular neurons. Deep layers of se_e3_e included [10,20,40] neurons at the hidden layers. The fitting neural network producing the force-field had [240, 240, 240] neurons on the hidden layers. The main part of the input-file for MLIP training is given below:

```
"model": {
    "type_map":    ["Fe", "B", "Si", "Ta"],
    "descriptor" :{
        "type": "hybrid",
        "list" : [ {
            "type":        "se_e2_a",
            "sel":         [150, 50, 50, 40],
            "rcut_smth":   2.0,
            "rcut":        7.0,
            "neuron":      [25, 50, 100],
            "resnet_dt":   false,
            "axis_neuron": 16,
            "seed":        16258,
        },
        {
            "type" : "se_e3",
            "sel":         [35, 20, 15, 15],
            "rcut_smth":   1.5,
            "rcut":        4.0,
            "neuron":      [10,20,40],
            "resnet_dt":   false,
            "seed":        23096,
            "_comment":    " that's all"
        }
        ]
    },
    "fitting_net" : {
        "neuron":     [240, 240, 240],
        "resnet_dt": true,
        "seed":       29621,
        "_comment":   " that's all"
    },
},
```

Developing a proper MLIP for a new system takes a lot of effort, especially for a multi-component system. For the training dataset, there should be many considerations.

We had about 30000 QMD (quantum molecular dynamics) frames used for investigation of the radial distribution functions in the melt. These frames we took as the initial training dataset (75% for training and 25% for validation) for MLIP. For example,
Fe382B76Si51Ta3, nfr=1833
Fe379Ta3Nb3B76Si51, nfr=4348
Fe380Ta5B76Si51, nfr=2510

Fe378B76Si50Ta8, nfr=22738
Fe380Ta5B76Si51, nfr4062
Fe388B74Si50, nfr=849
Fe385B76Si51, nfr=1026
…
Where "nfr" is the number of frames.

**The rest part of the training dataset was obtained using the different procedure with the help of DPGEN [2].**

Description of all the details of the DPGEN algorithm is a long and may be unnecessary story because the procedure we have used to build the training dataset mostly follows the DPGEN-paper [2] with the provided reference also in the body of the paper. So below we sketch only some particularly important points.

DPGEN is a software that allows automatic and highly optimised training of DEEPMD based MLIP if provided some initial training-validation dataset (in our case we take QMD data as this initial dataset). Usually, 4 replicas of DEEPMD MLIP are trained at the same time with independent random initialization of the neural networks before the start of the training. DPGEN typically runs 4 independent classical MD trajectories on LAMMPS (for each MLIP) and selects on fly "bad frames" for DFT calculations and final extension of the training dataset. Well trained MLIP should not strongly depend on the random initialization at the start of the training. So bad frames typically have large dispersion of forces evaluated by an ensemble average over 4 independent MLIP. "Large dispersion" criteria are flexible input parameters of DPGEN that users usually manually tune during different stages of DPGEN run [2]. Usually it implies, "much larger than the expected accuracy of MLIP compared to DFT".

DPGEN also contains the flexible engine to start automatically DEEPMD MLIP training, LAMMPS MD and a DFT (VASP in our case) on a supercomputer and process the results (e.g., extract energies, forces and virials and rebuild the training dataset).

We produced for the training dataset by DPGEN about 30000 frames with 512 atoms each. Classical LAMMPS MD simulations in DPGEN were performed using NPT Nosé–Hoover thermostat at pressure range 1 Pa – 10 GPa and temperature range 1000-2000 (K). The trajectory length of each LAMMPS MD run was 20000 steps with timestep=2 fs. For example, we investigated the following compositions
Fe378B76Si50Ta8, nfr=119,
Fe360B76Si51Ta25, nfr=242,
Fe309B76Si51Ta76, nfr=3798,
…

**To validate the performance of MLIP**, except for RDF and RMSE which were provided in the manuscript, we below include a graph showing the energies and forces along $x$, $y$ and $z$ directions from DFT based AIMD simulations and MLIP based classical MD simulations, see Fig. 2s.

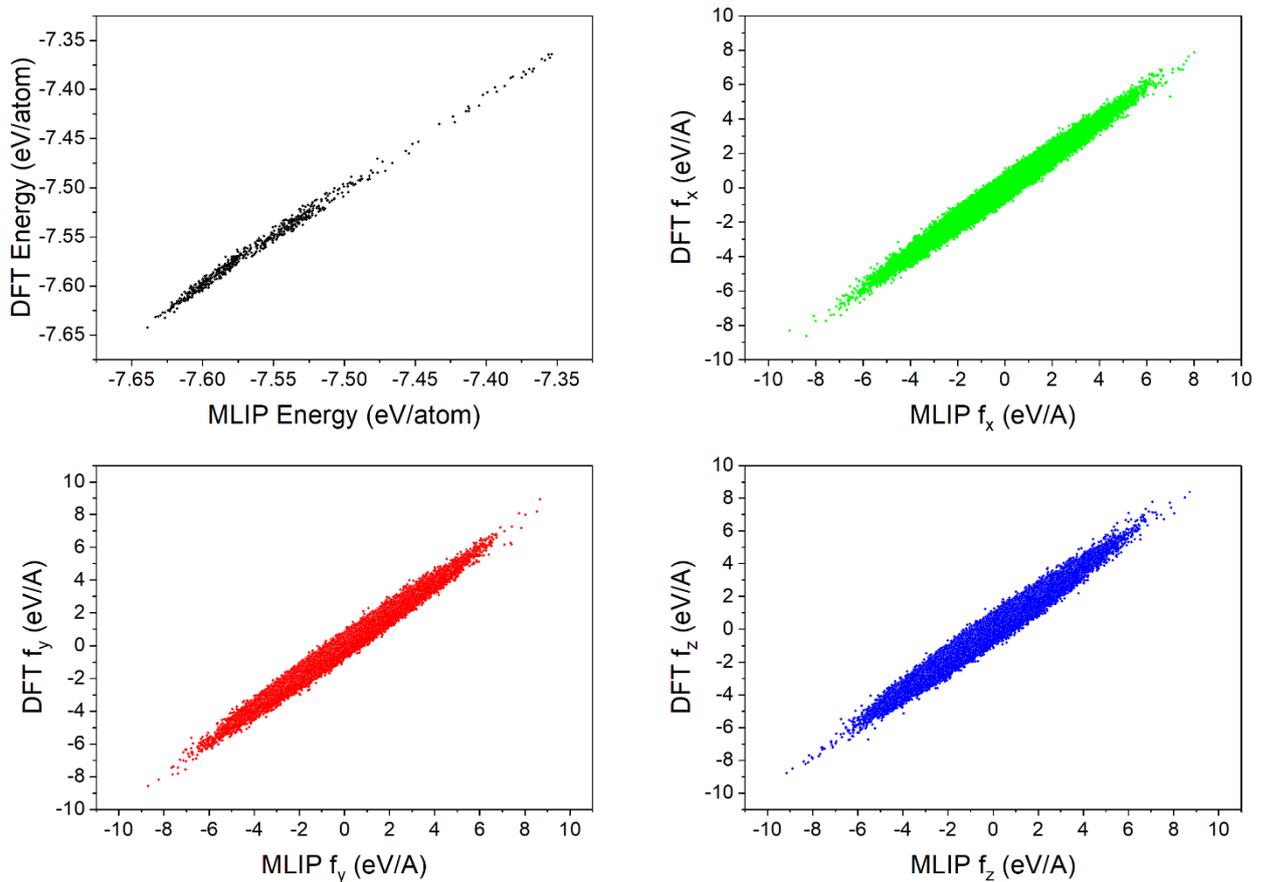

*Figure 2s. Accuracy of MLIP compared to DFT. The test is performed over all the training data.*

**About Cooling procedure**

For MD simulation, normally the melt-quenching method should be adopted to obtain the amorphous structures. In addition, more details below are included for MD simulation, such as the ensemble used, the cooling rate, the system size, ect..

Typical experimental rates of cooling $10^6 - 10^7$ K/s, which are 5 orders of magnitude lower than the effective cooling rate at the computer modeling. In classical MD with MLIP we had 500 particles and the cooling rate $7.5*10^{11}$ K/s. In QMD we had 512 atoms and the cooling rate $5*10^{13}$ K/s. In both cases the cooling was started from temperature T=2000 (K) and was performed using the *Nosé–Hoover* NPT thermostat. After the end of cooling, we switched on NVT *Nosé–Hoover thermostat to collect observables.*

**REFERENCES**

[1] H. Wang, L. Zhang, J. Han, E. Weinan, DeePMD-kit: A deep learning package for many-body potential energy representation and molecular dynamics, Computer Physics Communications. 228 (2018) 178-184. https://doi.org/10.1016/j.cpc.2018.03.016.

[2] Y. Zhang, H. Wang, W. Chen, J. Zeng, L. Zhang, H. Wang, E. Weinan, 2020. DPGEN: A concurrent learning platform for the generation of reliable deep learning based potential energy models, Computer Physics Communications. 253, 107206. https://doi.org/10.1016/j.cpc.2020.107206.